\documentclass[aps,pre,floatfix,nofootinbib,showpacs,twocolumn]{revtex4}
\usepackage{graphicx}
\usepackage{amsmath}
\usepackage{amssymb}

\renewcommand{\u}{\bar{u}}
\renewcommand{\v}{\bar{v}}

\newcommand{\BEQ}{\begin{eqnarray}}
\newcommand{\EEQ}{\end{eqnarray}}
\newcommand{\BEA}{\begin{eqnarray}}
\newcommand{\EEA}{\end{eqnarray}}

\renewcommand{\d}{{\rm d}}

\newcommand{\eps}{\varepsilon}

\newcommand{\tr}{{\rm tr}}

\renewcommand{\S}{{\rm\bf C}}
\newcommand{\R}{{\rm\bf H}}

\newcommand{\diag}{{\rm diag}}
\newcommand{\ssum}{{\sum}}
\renewcommand{\P}{{\bf P}}
\newcommand{\comment}[1]{}
\begin{document} 
\draft
\title{Optimal refrigerator}
\date{\today}
\author{ Armen E. Allahverdyan$^1$, Karen Hovhannisyan$^1$, Guenter Mahler$^2$}
\affiliation{$^1$Yerevan Physics Institute,
Alikhanian Brothers Street 2, Yerevan 375036, Armenia,\\
$^2$Institute of Theoretical Physics I, University of Stuttgart, 
Pfaffenwaldring 57, 70550 Stuttgart, Germany}

\begin{abstract} We study a refrigerator model which consists of two
$n$-level systems interacting via a pulsed external field. Each system
couples to its own thermal bath at temperatures $T_h$ and $T_c$,
respectively ($\theta\equiv T_c/T_h<1$). The refrigerator functions in
two steps: thermally isolated interaction between the systems driven by
the external field and isothermal relaxation back to equilibrium.  There
is a complementarity between the power of heat transfer from the cold
bath and the efficiency: the latter nullifies when the former is
maximized and {\it vice versa}. A reasonable compromise is achieved by
optimizing the product of the heat-power and efficiency over the
Hamiltonian of the two system. The efficiency is then found to be
bounded from below by $\zeta_{\rm CA}=\frac{1}{\sqrt{1-\theta}}-1$ (an
analogue of the Curzon-Ahlborn efficiency), besides being bound from
above by the Carnot efficiency $\zeta_{\rm C} = \frac{1}{1-\theta}-1$.
The lower bound is reached in the equilibrium limit $\theta\to 1$. The Carnot bound is
reached (for a finite power and a finite amount of heat transferred per
cycle) for $\ln n\gg 1$. If the above maximization is constrained by
assuming homogeneous energy spectra for both systems, the efficiency is
bounded from above by $\zeta_{\rm CA}$ and converges to it for $n\gg 1$.

\end{abstract}

\pacs{05.70.Ln, 05.30.-d, 07.20.Mc, 84.60.-h }

\comment{
Energy conversion, 84.60.-h

Irreversible thermodynamics, 05.70.Ln

Thermodynamics, 05.70.-a

Heat engines, 07.20.Pe

Refrigeration, 07.20.Mc

quantum statistical mechanics, 05.30.-d

}

\maketitle

\section{Introduction}

Thermodynamics studies principal limitations imposed on the performance
of thermal machines, be they macroscopic heat engines or refrigerators
\cite{lindblad,callen,landau}, or small devices in nanophysics \cite{q_t} and
biology \cite{venturi}.  Taking as an example a refrigerator driven by a
source of work, we recall three basic characteristics applicable to any
thermal machine:
\begin{itemize}

\item Heat $Q_c$ transferred per cycle of operation from a cold body
at temperature $T_c$ to a hot body at temperature $T_h$ ($T_h>T_c$). 

\item Power, which is the transferred heat $Q_c$ divided over the cycle
duration $\tau$.  

\item Efficiency (or performance coefficient) $\zeta=Q_c/W$, which
quantifies the useful output $Q_c$ over the work $W$ consumed from the
work-source for making the cycle. Note that work-consumption is
obligatory, since the heat is transferred from cold to hot, i.e.,
against its natural gradient. 
\end{itemize}

The second law imposes the Carnot bound 
\BEA
\zeta\leq \zeta_{\rm C} = T_c/(T_h-T_c) 
\nonumber
\EEA
on the efficiency of refrigeration \cite{callen}. Within the usual
thermodynamics the Carnot bound (both for heat-engines and
refrigerators) is reached only for a reversible, i.e., an infinitely
slow process, which means it is reached at zero power
\cite{callen,landau}. The practical value of the Carnot bound is
frequently questioned on this ground. 

The drawback of zero power is partially cured within finite-time
thermodynamics (FTT), which is still based on the quasi-equilibrium
concepts \cite{ftt}. For heat-engines FTT gives an {\it upper} bound
$\eta_{\rm opt}\leq \eta_{\rm CA}\equiv 1-\sqrt{T_c/T_h}$, where $\eta_{\rm opt}$ is the
efficiency at the maximal power of work-extraction \cite{ca}. Naturally,
$\eta_{\rm CA}$ is smaller than the Carnot upper bound $1-T_c/T_h$ for
heat-engines. 

Heat engines have recently been studied within
microscopic theories, where one is easily able to go beyond the
quasi-equilibrium regime
\cite{armen,tu,izumida_okuda,udo,esposito,jmod,domi,henrich}. For certain
classes of heat-engines the CA efficiency is a {\it lower} bound for the
efficiency at the maximal power of work \cite{armen,tu,izumida_okuda}.
This bound is reached at the quasi-equilibrium situation $T_h\to T_c$ in
agreement with the finding of FTT. The result is consistent with other
studies \cite{udo,esposito}. 

The interest in small-scale refrigerators is triggered by the importance
of cooling processes for functioning of small devices and for displaying
quantum features of matter \cite{q_t,henrich,kosloff_jap,feldman,kosloff_minimal_temperature,segal,rezek}.
In particular, the theory of these refrigerators can provide answers to
several basic questions such as how the third law limits the performance
of a cooling machine at low temperatures \cite{kosloff_jap}, and how
small are the temperatures reachable within a finite working time and
under a reasonable amount of resource. Naturally, the small-scale
refrigerators should also operate at a finite power. Note that the
mirror symmetry between heat-engines and refrigerators, which is
well-known for the zero-power case \cite{callen}, does not hold
more generally \cite{yan_chen}. 

The present situation with finite-power refrigerators is somewhat unclear
\cite{yan_chen,velasco,jimenez,unified}. Here maximizing the power of
cooling does not lead to reasonable results, since there is an
additional complementarity (not present for heat engines)
\cite{yan_chen,velasco,feldman,kosloff_minimal_temperature}: when maximizing the heat-transfer power
one simultaneously minimizes the efficiency to zero, and {\it vice
versa}.

Here we intend to study optimal regimes of finite-power refrigeration
via a model which can be optimized over almost all of its parameters.
The model represents a junction immersed between two thermal baths at
different temperatures and driven by an external work-source.  This type
of models is frequently studied for modelling heat transport; see, e.g.,
\cite{q_t,segal,dhar}. Our model is quantum, but it admits a classical
interpretation, because all the involved density matrices will be
diagonal [in the energy representation] at initial and final moments of
studied processes \footnote{This aspect is similar to the Ising model.
This is a model for quantum-mechanical
spin-$\frac{1}{2}$, but it can be given a classical
interpretation via an overdamped particle moving in a asymmetric
double-well potential. If the transversal components of the quantum spin
are excited, this analogy breaks down. However, it still holds for the
spin-flip process, where the transversal components are absent both
initially and finally. In fact, the dynamics of the Ising model is 
introduced via such spin-flip processes, and this dynamics admits a 
classical interpretation.}.

This paper is organized as follows. The model is introduced in section
\ref{model}. Here we also show that the efficiency of the model is
bounded by the Carnot value, and provide a general discussion of the
refrigeration power.  We confirm the heat-power-efficiency
complementarity in section \ref{compl} and conclude that the most
meaningful way of optimizing its functioning is to maximize the product
of efficiency and the heat power. The optimization procedure is reported
in section \ref{max_product}. We discuss the quasi-equilibrium limit of
our model in section \ref{quasi}. There we show that there is a {\it
lower} bound $\zeta_{\rm CA}= -1+1/\sqrt{1-\theta}$ ($\theta\equiv
T_c/T_h$) for the efficiency, in addition to the upper Carnot bound
$\zeta_{\rm C} = \frac{\theta}{1-\theta}$.  The same expression
$\zeta_{\rm CA}$ was obtained within finite-time thermodynamics as an
{\it upper} bound when optimizing the product of heat-power and
efficiency or the ratio of the efficiency over the cycle time
\cite{yan_chen,velasco}. Section \ref{carnot_richting} discusses the
attainability of the Carnot efficiency at a finite power.  Entropy
production inherent in the functioning of the model refrigerator is
studied in section \ref{entropy_production}, while in section
\ref{class} we outline consequences of constraining features of the
model to the quasi-classical domain. This constraint allows to reproduce
the prediction of FTT on the upper bound of $\zeta_{\rm CA}$. We
summarize in section \ref{summa_contra_gentiles}. Some technical
questions are relegated to Appendix. 

\section{The model}
\label{model}

\begin{figure}
\includegraphics[width=8.5cm]{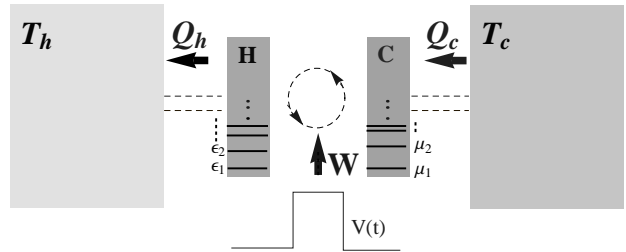}
\caption{ The refrigerator model. Two systems $\R$ 
and $\S$ operate between two baths at temperatures
$T_c<T_h$ and are driven by an external potential $V(t)$. 
$W$ and $Q_c$ and $Q_h$ are, respectively, the work put into
the overall system and the heats transferred from the cold bath and to the hot bath. } 
\label{f1}
\end{figure}

Consider two quantum systems $\R$ and $\S$ with Hamiltonians $H_\R$
and $H_\S$, respectively. Each system has $n$ energy levels. 
$\R$ and $\S$ constitute the working medium of our refrigerator; 
see Fig.~\ref{f1}.

Initially, $\R$ and $\S$ do not interact and are in equilibrium
at temperatures $T_h=1/\beta_h>T_c=1/\beta_c$ [we set $k_{\rm B}=1$]:
\BEA
\label{1}
\rho={e^{-\beta_h H_\R}}/{\tr\, [e^{-\beta_h H_\R}]}, ~~
\sigma={e^{-\beta_c H_\S}}/{\tr\, [e^{-\beta_c H_\S}]}, 
\EEA
where $\rho$ and $\sigma$ are the initial Gibbsian density matrices of $\R$ and $\S$, respectively. 
We write
\begin{gather}
\label{2}
\rho=\diag [r_n,...,r_1], ~~~~~~~~~~ 
\sigma=\diag [s_n,...,s_1], \\
H_\R=\diag [\eps_n,...,\eps_1=0\,],~~
H_\S=\diag [\mu_n,...,\mu_1=0\,], 
\label{2star}
\end{gather}
where $\diag[a,..,b]$ is a diagonal matrix with entries $(a,...,b)$, and
where without loss of generality we have nullified the lowest energy level
of both $\R$ and $\S$.  Thus the overall initial density matrix is 
\BEA
\label{101}
\Omega_{\rm in}=\rho\otimes\sigma, 
\EEA
and the initial Hamiltonian 
$H_\R\otimes 1+1\otimes H_\S $. 

The goal of any refrigerator is to transfer heat from the cooler bath to
the hotter one at the expense of consuming work from an external source.
The present refrigerator model functions in two steps: thermally isolated 
work-consumption and isothermal relaxation; 
see Fig.~\ref{f1}. Let us describe these steps in detail.

{\bf 1.} $\R$ and $\S$ interact with each other and with the external
sources of work. The overall interaction is described via a
time-dependent potential $V(t,\delta)$ in the total Hamiltonian 
\BEA
\label{bobo}
H(t,\delta)=H_\R\otimes 1+1\otimes H_\S+V(t,\delta) 
\EEA
of $\R+\S$. The interaction process is thermally isolated: $V(t,\delta)$ is
non-zero only in a short time-window $0\leq t\leq \delta$ and is so
large there that the influence of all other couplings [e.g., couplings
to the baths] can be neglected [pulsed regime]. The
time-dependent potential $V(t,\delta)$ may explicitly depend on the
coupling time $\delta$.

Thus the dynamics of $\R+\S$ is unitary for $0\leq t\leq \delta$: 
\BEA
\Omega_{\rm f}\equiv \Omega (\delta )={\cal U}\,
\Omega_{\rm i}\, {\cal U}^\dagger, ~~{\cal U}={\cal T}e^{-\frac{i}{\hbar} \int_0^\delta \d s V(s,\delta)  },
\label{blum}
\EEA
where $\Omega_{\rm i}=\Omega(0)=\rho\otimes\sigma$ is the initial state defined in
(\ref{1}), $\Omega_{\rm f}$ is the final density matrix, ${\cal
U}$ is the unitary evolution operator, and where ${\cal T}$ is the time-ordering operator. 
The work put into $\R+\S$ reads \cite{lindblad,callen}
\BEA
W=E_{\rm f}-E_{\rm i}=\tr [\,( H_\R\otimes 1+1\otimes H_\S)\, (\Omega_{\rm f}-\Omega_{\rm i}) \,],
\label{work}
\EEA
where $E_{\rm f}$ and $E_{\rm i}$ are initial and final energies of $\R+\S$.

{\bf 2.} Once the overall system $\R+\S$ arrives at the final state
$\Omega_{\rm fin}$, $V(t,\delta)$ is switched off, and $\R$ and $\S$ (within
some relaxation time) return back to their initial states (\ref{1})
under influence of the hot and cold thermal baths, respectively.  Thus
the cycle is complete and can be repeated again.  Because the energy is
conserved during the relaxation, the hot bath gets an amount of heat
$Q_h$, while the cold bath gives up the amount of heat $Q_c$:
\BEA
\label{heats}
Q_h=\tr (H_\R[\,{\rm tr}_\S\Omega_{\rm f}-\rho]),\, \, \,\,
Q_c=\tr (H_\S[\sigma-{\rm tr}_\R\Omega_{\rm f}]),
\EEA
where ${\rm tr}_\R$ and ${\rm tr}_\S$ are the partial traces.
Eq.~(\ref{1}) and the unitarity of ${\cal U}$ lead to
\BEA 
\label{43}
\beta_h Q_h-\beta_c Q_c=S(\Omega_{\rm f}||\Omega_{\rm i})\equiv {\rm tr}[\Omega_{\rm f}\ln \Omega_{\rm f}
-\Omega_{\rm f}\ln \Omega_{\rm i}],
\EEA 
where $S(\Omega_{\rm f}||\Omega_{\rm i})\geq 0$ is the relative entropy, which employed 
in deriving thermodynamic bounds since \cite{lindblad,partovi}.

$S(\Omega_{\rm f}||\Omega_{\rm i})$ nullifies if and only if
$\Omega_{\rm f}=\Omega_{\rm i}$; otherwise it is positive.
Eq.~(\ref{43}) is the Clausius inequality, with $S(\Omega_{\rm
f}||\Omega_{\rm i})\geq 0$ quantifying the entropy production. This
point will be re-addressed and confirmed in section
\ref{entropy_production}. 

Eqs.~(\ref{work}--\ref{43}) and the energy conservation $Q_h=W+Q_c$
imply 
\BEA
\label{otdel}
(\beta_c-\beta_h)Q_c\leq\beta_h W, 
\EEA
meaning that in the
refrigeration regime we have $Q_c>0$ and thus $W>0$. Thus within the step
{\bf 1} the work source transfers some energy from $\S$ to $\R$, while
in the step {\bf 2}, $\S$ recovers this energy from the cold bath thereby
cooling it and closing the cycle. 

Eq.~(\ref{43}) leads to the Carnot bound for the
efficiency $\zeta$ [we denote $\theta\,\equiv\, T_c/T_h<1$]
\BEA
\label{laplace}
\zeta\,\equiv\, \frac{Q_c}{W}
\,= \,\frac{\theta}{1-\theta}\,-\,\frac{S(\Omega_{\rm f}||\Omega_{\rm i})}{(\beta_c-\beta_h)W}
\,\leq\, \frac{\theta}{1-\theta}\,\equiv\,\zeta_{\rm C}. 
\EEA
We note from (\ref{laplace}) that the deviation from the Carnot bound is controlled by the ratio of the 
entropy production $S(\Omega_{\rm f}||\Omega_{\rm i})$ to the work $W$.

We note in passing that all quantities introduced so far are meaningful
also without the stage {\bf 2}. Then the problem reduces to cooling the
initially equilibrium system $\S$ with help of the work-source and the
system $\R$. Both the work-source and $\R$ are clearly necessary to
achieve cooling \footnote{Indeed, if $\S$ and the work-source form a closed system, no cooling
is possible due to the Thomson's formulation of the second law \cite{jmod} (cyclic 
processes cannot lead to work-extraction). If
$\R$ and $\S$ form a closed system, then $W=0$ and
no cooling is possible due to (\ref{otdel}).}. $Q_c$ quantifies the amount of cooling, while
$\zeta$ accounts for the relative effort of cooling. 

\subsection{Power}
\label{po}

Recall that the power of refrigeration $Q_c/\tau$ is defined as the
ratio of the transferred heat $Q_c$ to the cycle duration $\tau$. For
our model $\tau$ is limited mainly by the duration of the second stage,
i.e., $\tau$ should be larger than the relaxation time $\tau_{\rm rel}$,
which depends on the concrete physics of the system-bath coupling. 

Though some aspects of the following discussion are rather general, 
it will be useful to have in mind a concrete relaxation scenario. 
Consider the collisional relaxation scenario, where the
target system interacts with independent bath particles via successive
collisions; see \cite{partovi,mityugov} and Appendix
\ref{app_collisions}.  For our purpose the target system is $\R$ or $\S$
that interact with, respectively, the hot and cold bath.  Each collision
lasts a time $\tau_{\rm col}$, which is much smaller than the
characteristic time $\tau_{\rm btw}$ between two collisions. The
interaction Hamiltonian between the target system and a bath particle is
conserved, so that no work is done in switching the system-bath
interaction on and off; see \cite{partovi,mityugov} and Appendix
\ref{app_collisions_0}.  

The relaxation process is typically (but not always) exponential with
the characteristic relaxation time depending on the collisional
interaction; see Appendix \ref{app_collisions_1}. This time can be much
smaller than any characteristic time of $\R$ or $\S$. Since the two
baths act on $\R$ and $\S$ independently, the overall relaxation process
drives $\R+\S$ to the initial state (\ref{1}, \ref{101}). 

If the interaction time $\delta$ of $V(t,\delta)$ [see (\ref{blum})] is
also much smaller than $\tau_{\rm btw}$, one realizes a
thermally-isolated process, because the overlap between the pulse and a
collision can be neglected \footnote{ Analogous conclusion on the
irrelevance of the system-bath interaction during the pulse action is
obtained when this interaction is always on, but its magnitude is small
[weak-coupling]. Now the relaxation time is much larger than the
internal characteristic time of $\R$ and $\S$.  Because the system-bath
interaction is always on, there will be a contribution in the work
(\ref{work}) coming from the system-bath interaction Hamiltonian
\cite{ruben}. This contribution arises even when the conditions for the
pulsed regime hold \cite{ruben}. However, within the weak coupling
assumption this additional contribution is proportional to the square of
the system-bath interaction constant and can be neglected \cite{ruben}.
We stress that this additional contribution does not arise within the
collisional relaxation scenario, because the pulse and collisions are
well-separated in time. }.

To achieve a cyclic process within the exponential relaxation with the
relaxation time $\tau_{\rm rel}$, the cycle time $\tau$ should be larger
than $\tau_{\rm rel}$. For each cycle the deviation of the
post-relaxation state from the exact equilibrium state (\ref{1}, \ref{101}) 
will be of order $e^{-\tau/\tau_{\rm rel}}$.  Thus if
the ratio $\tau/\tau_{\rm rel}$ is simply large, but finite, one can
perform roughly $\sim e^{\tau/\tau_{\rm rel}}\gg 1$ number of cycles at
a finite power, before deviations from cyclicity would accumulate and 
the refrigerator will need resetting. 

Though, as we stressed above, the relaxation process is normally
exponential \footnote{ More generally, the relaxation need not be
exponential, but it still can be such that although the difference
between the system density-matrix at time $t$ and the corresponding
Gibbsian density matrix goes to zero for $t\to\infty$, this difference
does not turn to zero after any finite $t$.  One of referees of this
paper pointed out to us that {\it i)} the latter feature holds for a
rather general class of relaxation processes taking place under a
constant Hamiltonian; {\it ii)} it is rooted in the
Kubo-Martin-Schwinger (KMS) condition \cite{kms} for correlation
function evaluated over an equilibrium state; see \cite{landau} for a
heuristic version of this argument; {\it iii)} the collisional
relaxation is different in this respect, because the Hamiltonian is not
constant. As we stress in Appendix \ref{app_collisions}, a general 
point of the collisional relaxation is that no work is
involved in this time-dependence.}, there are also situations within the
collision relaxation scenario, where the system settles in the
equilibrium state after just one inter-collision time $\tau_{\rm btw}$;
see section \ref{popo} and Appendix \ref{app_collisions_1} for details.
The above limitations on the number of cycles does not apply to this
relaxation scenario.

\subsubsection{Comparing with the power of the Carnot cycle}
\label{carcar}

The above situation does differ from the power consideration of usual
(reversible) thermodynamic cycles, e.g., the Carnot cycle
\cite{callen,landau,sekimoto,sekimoto_sasa}. There the external fields
driving the working medium through various stages have to be much slower
than the relaxation to the momentary equilibrium. The latter means that
the working medium is described by its equilibrium Gibbs distribution
with time-dependent parameters. The condition of momentary equilibrium
for the working medium is necessary for the Carnot cycle to reach the
Carnot efficiency \cite{callen,landau}.

The precise meaning of the external fields being slow is important
here. If $\tau_{\rm F}$ is the characteristic time of the fields, then
the deviations from the momentary equilibrium are of order ${\cal
O}[\frac{\tau_{\rm rel}}{\tau_{\rm F}}]$
\cite{landau,sekimoto,sekimoto_sasa}. This fact is rather general and
does not depend on details of the system and of the studied process,
e.g., it does not depend whether the process is thermally isolated or
adiabatic \footnote{If a slow thermally isolated process is performed on
a finite system, there are additional limitations in achieving the
momentary equilibrium; see \cite{sekimoto,minima} for more details.
These limitations are however not essential for the present argument.}. 
In particular, it is this deviation of the state from the momentary equilibrium that
brings in the entropy production (or work dissipation) of order of 
${\cal O}[\left(\frac{\tau_{\rm rel}}{\tau_{\rm F}}\right)^2]$
\cite{landau,sekimoto,sekimoto_sasa}.

Thus performing the reversible Carnot cycle with (approximately) the
Carnot efficiency means keeping the ratio $\frac{\tau_{\rm
rel}}{\tau_{\rm F}}$ very small. 

Now there are two basic differences between the Carnot cycle and our situation: 

\begin{itemize}
\item In our case we do not require the working medium to be close to its
momentary equilibrium state during the whole process. It suffices that
the medium gets enough time to relax to its final equilibrium. 

\item A small, but finite $\frac{\tau_{\rm rel}}{\tau_{\rm F}}$ for
the Carnot cycle situation means that deviations from the momentary
equilibrium are visible already within one cycle. In contrast, a small,
but finite $\frac{\tau_{\rm rel}}{\tau}$ for our situation means that we
can perform an exponentially large number of cycles before deviations
from the cyclicity will be sizable. Here is a numerical example. Assume
that $\frac{\tau_{\rm rel}}{\tau}=\frac{\tau_{\rm rel}}{\tau_{\rm
F}}={1}/{20}$. For the standard Carnot cycle already within one cycle
the deviation from the momentary equilibrium will amount to $0.05$. In
our situation the same amount $e^{-3}=0.0498$ of deviation from the
cyclicity will come after $e^{17}=2.4\times 10^7$ cycles. This is a
large number, especially taking into account that no realistic machine
is supposed to work indefinitely long. Such machines do need resetting
or repairing. The point is that our machine can perform {\it many}
cycles at a finite power before any resetting is necessary. 

\end{itemize}

\section{Complementarity between the transferred heat and efficiency}
\label{compl}

We now proceed to optimizing the functioning of the refrigerator over
the three sets of available parameters: the energy spacings
$\{\eps_k\}_{k=2}^n$, $\{\mu_k\}_{k=2}^n$, and the unitary operators
${\cal U}$. It should be evident from (\ref{bobo}, \ref{1}) that
optimizing over these parameters is equivalent to optimizing over the
full time-dependent Hamiltonian $H(t,\delta)$ of $\R+\S$. We stress in this
context that no limitations on the magnitude of $V(t,\delta)$ are imposed. This
means that the unitary operator can in principle be generated in an
arbitrary short coupling time $\delta$. 
  
We start by maximizing the transferred heat $Q_c=\tr (H_\S[\sigma-{\rm
tr}_\R\Omega_{\rm f}])$, which is the main characteristics of the
refrigerator. Since $\tr [H_\S\sigma]$ depends only on
$\{\mu_k\}_{k=2}^n$, we choose $\{\eps_k\}_{k=2}^n$ and $V(t)$ so that
the final energy $\tr [H_\S\Omega_{\rm f}]$ attains its minimal value
zero. Then we maximize $\tr [H_\S\sigma]$ over $\{\mu_k\}_{k=2}^n$.
Note from (\ref{2})
\BEA
1\otimes H_\S &=& {\rm diag}[\,\mu_1\,\,,\ldots,\,\,\,\mu_1,\ldots,
                             \,\,\mu_n\,\,\,,\ldots,\,\,\mu_n\,\, ],\nonumber\\
\Omega_{\rm i}=\rho\otimes\sigma &=& {\rm
  diag}[\, s_1r_1,\ldots,s_1r_n,\ldots,
                            s_nr_1,\ldots,s_nr_n\, ].\nonumber
\EEA
It is clear that $\tr [H_\S\Omega_{\rm f}]=\tr [H_\S{\cal U}\Omega_{\rm
 i}{\cal U}^\dagger]$ goes to zero when, e.g., $r_2=\ldots=r_n\to 0$ 
($\eps\equiv\eps_2=\ldots=\eps_n\to \infty$),
while ${\cal U}$ amounts to the SWAP operation 
${\cal U}\rho\otimes\sigma{\cal U}^\dagger = \sigma\otimes\rho$. 
It is checked by a direct
inspection that the maximization of the initial energy 
$\tr [H_\S\sigma]$ over $\{\mu_k\}_{k=2}^n$ produces the same
structure of $n-1$ times degenerate upper energy levels
$\mu\equiv\mu_2=\ldots=\mu_n$. Denoting 
\begin{gather}
\label{burundi}
v\equiv s_2=..=s_n= e^{-\beta_c \mu},~~~ u\equiv r_2=..=r_n= e^{-\beta_h \eps},
\end{gather}
we obtain for $Q_c$
\BEA
\label{10} 
Q_c=
T_c \ln\left[\frac{1}{v}\right]\,\frac{(v-u)(n-1) }{[\,1+(n-1)v\,][\,1+(n-1)u\,]}, 
\EEA
where according to the above discussion, $Q_c$ is maximized for $u\to
0$, and where $v$ is to be found from maximizing $Q_c|_{u\to 0}$ in 
(\ref{10}) over $v$, i.e., $v$ is
determined via 
\BEA
1+(n-1)v+\ln v=0. 
\EEA
For the efficiency we get for the present situation ($\R$ and $\S$
have $n-1$ times degenerate upper levels, while ${\cal U}$ amounts to
the SWAP operation):
\BEA \label{11}
\zeta\,=\,\frac{Q_c}{W}\,= \frac{\mu }{\eps-\mu}
=\frac{\,\theta\, \ln[\,\frac{1}{v}\,]}{ \ln[\,\frac{1}{u}\,]-\theta\ln[\,\frac{1}{v}\,]}.
\label{peshavar}
\EEA
The maximization of $Q_c$ led us to $u\to 0$, which then means that
$\zeta$ in (\ref{11}) goes to zero.  

Thus $\S$ can be cooled down to its ground state ($\tr [H_\S\Omega_{\rm
f}]\to 0$), but at a vanishing efficiency, i.e., at expense of an
infinite work. To make this result consistent with the classic message of the
third law \cite{klein}, we should slightly adjust the latter: one cannot
reach the zero temperature [of an initially equilibrium system] in a
finite time {\it and} with finite resources [infinite work is not a
finite resource]. At any rate, one should note that the classic formulation of the third law motivates its
operational statement using exclusively equilibrium concepts. 
Modern perspectives on the third law are discussed in 
\cite{kosloff_jap,kosloff_minimal_temperature,rezek,grif,scully}.

Note that the efficiency $\zeta$ in (\ref{11})
reaches its maximal Carnot value $\theta/(1-\theta)$ for 
\BEA
\label{bekum}
u=v, 
\EEA
which
nullifies the transferred heat $Q_c$; see (\ref{10}). 

Now we have to show
that $Q_c$ tends to zero
upon maximizing $\zeta$ over {\it all} free parameters
$\{\eps_k\}_{k=2}^n$, $\{\mu_k\}_{k=2}^n$ and ${\cal U}$.
Denoting $\{| i_\R\rangle \}_{k=1}^n$ and $\{| i_\S\rangle \}_{k=1}^n$
for the eigenvectors of $H_\R$ and $H_\S$, respectively, we note 
from (\ref{work}, \ref{heats}) 
that $W$ and $Q_c$ feel ${\cal U}$ only via the matrix
\BEA
\label{gomel}
C_{ij\,|\,kl}=|\langle i_\R j_\S |{\cal U}| k_\R l_\S\rangle|^2.
\EEA
This matrix is double-stochastic \cite{olkin}: 
\BEA
\label{chernigov}
{\ssum}_{ij }C_{ij\,|\,kl}={\ssum}_{kl }C_{ij\,|\,kl}=1.
\EEA
Conversely, for any double-stochastic matrix $C_{ij\,|\,kl}$ there
is some unitary matrix $U$ with matrix elements $U_{ij\,|\,kl}$,
so that $C_{ij\,|\,kl}=|U_{ij\,|\,kl}|^2$ \cite{olkin}. Thus, when
maximizing various functions of $W$ and $Q_c$ over the unitary ${\cal
U}$, we can directly maximize over the $(n^2-1)^2$ independent elements
of $n^2\times n^2$ double stochastic matrix $C_{ij\,|\,kl}$.  

We did not find an analytic way of carrying out the complete
maximization of $\zeta$ over all free parameters. Thus we had to rely on
numerical recipes of Mathematica 7, which for $n=1,\ldots, 5$ confirmed
that $Q_c$ nullifies whenever $\zeta$ reaches (along any path) its
maximal Carnot value. We believe this holds for an arbitrary $n$, though
we lack any rigorous proof of this assertion.

\begin{figure}
\includegraphics[width=8cm]{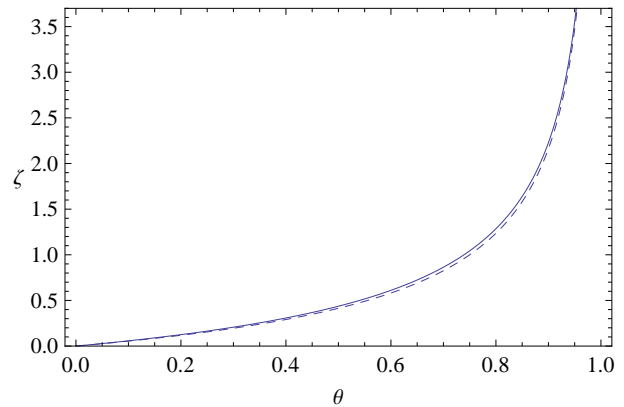}
\caption{ Solid line: efficiency $\zeta$ of the optimized refrigerator versus the temperature ratio $\theta=
T_c/T_h$ for $n=3$; see (\ref{11}). In the scale of this figure $\zeta(n=2)$ and $\zeta(n=3)$ are almost indistinguishable.
Dashed line: the lower bound $\frac{1}{\sqrt{1-\theta}}-1$. } 
\label{f2}
\end{figure}

\section{Maximizing the product of the transferred heat and efficiency.}
\label{max_product}

We saw above that neither $Q_c$ nor $\zeta$ are good target quantities for determining
an optimal regime of refrigeration. But 
\BEA
\chi \equiv Q_c\zeta,
\EEA
is such a target quantity, as will be seen shortly. This is the most
natural choice for our setup. This choice was also employed in
\cite{yan_chen}.
Refs.~\cite{unified,feldman,kosloff_minimal_temperature} report on
different approaches to defining refrigeration regimes. 

The numerical maximization of $\chi=\zeta Q_c$ over
$\{\eps_k\}_{k=2}^n$, $\{\mu_k\}_{k=2}^n$ and ${\cal U}$ has been
carried out for $n=1,\ldots,5$ along the lines discussed around
(\ref{gomel}, \ref{chernigov}). It produced the same structure: both
$\R$ and $\S$ have $n-1$ times degenerate upper levels, see
(\ref{burundi}), and the optimal ${\cal U}$ again corresponds to SWAP
operation \footnote{Let us recall how the SWAP is defined via a pure-state base. 
Let $\{ |k\rangle_1\}_{k=1}^n$ be an orthonormal base in the Hilbert
where $\rho$ lives.
Let also $\{ |k\rangle_2\}_{k=1}^n$ be an orthonormal base in the Hilbert
space where $\sigma$ lives. 
Any unitary operator acting on the composite Hilbert space can be 
defined with respect to the orthonormal base $|k\rangle_1\otimes |l\rangle_2$, where $k,l=1,\ldots,n$.
Let us now define for all pairs $k$ and $l$:
$U_{\rm SWAP} |k\rangle_1\otimes |l\rangle_2 =|l\rangle_1\otimes |k\rangle_2$.}: 
\BEA
\label{cobra}
\Omega_{\rm i}=\rho\otimes\sigma,~~~~~~ \Omega_{\rm f}= \sigma\otimes \rho.
\EEA

Recalling the expression (\ref{bobo}) for the total Hamiltonian
$H(t,\delta)$, we can state this result as follows: there exist a
coupling potential $V(t,\delta)$ that for a given coupling time $\delta$
generates the unitary SWAP operation following to (\ref{blum}). This
operation does not explicitly depend on $\delta$, because $V(t,\delta)$
depends on $\delta$; see also our discussion in the beginning of section
\ref{compl}. Note that both the initial and final states in
(\ref{cobra}) are diagonal in the energy representation. Evidently, the
intermediate state $\Omega(t)$ for $0<t<\delta$ is not diagonal in this
representation. 

The efficiency $\zeta$ and the transferred heat $Q_c$ are given by, respectively, 
(\ref{10}) and (\ref{11}), where instead of $u$ and $v$ we should substitute
$\u$ and $\v$, respectively. The latter two quantities are obtained from maximizing 
$\chi=\zeta Q_c$, 
\BEA
\label{gyurza}
\label{14}
\chi(\u,\v)= 
\frac{T_c\theta (n-1)(\v-\u)\ln^2\frac{1}{\v}}
{[\ln\frac{1}{\u}-\theta\ln\frac{1}{\v}][1+(n-1)\u][1+(n-1)\v]},
\EEA
where $\u$ and $\v$ are found from maximizing $\chi(u,v)$ via
$\partial_u \chi=\partial_v \chi=0$. Note that $\u$ and $\v$ depend on
$\theta=T_c/T_h$. The efficiency $\zeta$ and the transferred heat $Q_c$
are given, respectively, by (\ref{11}) and (\ref{10}) with $u\to\u$ and
$v\to\v$. 

Though we have numerically checked these results for $n\leq 5$ only, we
again trust that they hold for an arbitrary $n$ (one can, of course,
always consider the above structure of energy spacings and ${\cal U}$ as
a useful ansatz).

SWAP is one of the basic gates of quantum information processing
\cite{galindo}; see \cite{domi} for an interesting discussion on the
computational power of thermodynamic processes. SWAP is sometimes
realized as a composition of more elementary unitary operations, but its
direct realizations in realistic systems also attracted attention; see,
e.g., \cite{swap} for a direct implementation of SWAP in quantum optics.
Note that for implementing the SWAP as in (\ref{cobra}) the external
agent need not have any information on the actual density matrices
$\rho$ and $\sigma$. 

\subsection{Effective temperatures}

Since the state $\Omega_{\rm f}$ of $\R+\S$ after the action of $V(t)$ is
$\sigma\otimes\rho$, and because in the optimal regime the upper level
for both $\R$ and $\S$ is $n-1$ times degenerate, one can introduce
non-equilibrium temperatures $T_h'$ and $T_c'$ for respectively $\R$ and
$\S$ via [note (\ref{1})]
\BEA
\sigma={e^{-\beta_h' H_\R}}/{\tr\, [e^{-\beta_h' H_\R}]}, ~~
\rho={e^{-\beta_c' H_\S}}/{\tr\, [e^{-\beta_c' H_\S}]}, 
\EEA
where we recall that $\sigma$ ($\rho$) is the state of $\R$ ($\S$) after 
applying the pulse. Using (\ref{burundi}) we deduce
\BEA
\label{aharon1}
T_h'=T_h\,\frac{\ln\frac{1}{\u}}{\ln\frac{1}{\v}}, ~~~~
T_c'=T_c\,\frac{\ln\frac{1}{\v}}{\ln\frac{1}{\u}},
\EEA
where
$\v=e^{-\beta_c\bar{\mu}}$ and $\u=e^{-\beta_h\bar{\eps}}$; see
(\ref{burundi}).  This implies 
\BEA
\label{aharon2}
T_cT_h=T'_cT'_h.  
\EEA
As expected, the
refrigeration condition $\v>\u$, see (\ref{10}, \ref{14}), is equivalent
to 
\BEA
\label{aharon3}
T_c'<T_c<T_h<T_h', 
\EEA
i.e., after the pulse the cold system gets colder, while the hot system
gets hotter.  Note that the existence of temperatures $T'_c$ and $T'_h$
was not imposed, they emerged out of optimization. In terms of these
temperatures the efficiency (\ref{peshavar}) is conveniently written as
\BEA
\label{kandagar}
\zeta = \frac{T_c}{T_h'-T_c} = \frac{T_c'}{T_h-T_c'}.
\EEA

We eventually focus on two important limits: 
quasi-equilibrium $\theta\to 1$, and the regime $\ln n\gg 1$. 

\section{Quasi-equilibrium regime $\theta\to 1$: a lower bound for the efficiency}
\label{quasi}

In this regime the temperatures $T_h$ and $T_c$ are nearly equal to each
other: $\theta\equiv T_c/T_h\to 1$. 

First we note that sharply at $\theta=1$, $\chi$ reads
\BEA
\label{15}
\chi(a)|_{\theta=1}
= T_c \,\theta (n-1)\,
{[1+(n-1)a]^{-2}}\,a\, \ln^2{a},
\EEA
where
\BEA
\u=\v=a, \nonumber
\EEA
and where $a$ is given by $\partial_a \chi(a)|_{\theta=1}=0$:
\BEA
\label{golem}
[{(n-1)a-1}]\, \ln a=2 [{(n-1)a+1}].
\EEA
We now work out the optimal $\u$ and $\v$ for $\theta\to 1$. It can be seen from (\ref{14})
that the proper expansion parameter for $\theta\to 1$ is $x\equiv \sqrt{1-\theta}$.
We write
\BEA
\u=a+{\ssum}_{k=1}a_kx^k,~~\v=a+{\ssum}_{k=1}(a_k+b_{k-1})x^k.\,
\label{lahabana}
\EEA
We substitute (\ref{lahabana}) into $\partial_u \chi=0$ and $\partial_v
\chi=0$ and expand them over $x$. Both expansions start from terms of
order ${\cal O}(x^0)$. Now $a_k$ and $b_k$ are determined by equating
to zero the ${\cal O}(x^k)$ terms in $\partial_u \chi=0$ and $\partial_v
\chi=0$. Thus the ${\cal O}(x^0)$ terms together with (\ref{golem})
define $b_0$:
\BEA
b_0=a\ln\frac{1}{a},
\label{makich}
\EEA
which should be non-negative due to $v>u$. The ${\cal O}(x^1)$ terms together with (\ref{golem}) and (\ref{makich})
define $a_1$ and $b_1$:
\BEA
a_1=-\frac{a}{2}\ln\frac{1}{a}, \quad b_1=-\frac{a[24+\ln^2 a]}{48}\ln\frac{1}{a},
\label{hskich}
\EEA
and so on. 
Eqs.~(\ref{lahabana}, \ref{makich}, \ref{hskich}) imply for the efficiency at $\theta\to 1$ ($x=\sqrt{1-\theta}$)
\begin{gather}
\zeta= \frac{1}{x}-1+\frac{\ln^2 a}{48}-\frac{[48+\ln^2a]\ln^2 a}{1536}\,x
+{\cal O}(x^2).
\label{capo}
\end{gather}
Note that the expansion (\ref{capo}) does not apply for $ n\to\infty$,
since in this limit $a(n-1)\simeq 1-\frac{4}{\ln [n-1]}$; see
(\ref{golem}). Thus, in the limit $\theta\to 1$, 
$Q_c$ scales as $\propto \sqrt{1-\theta}$, 
\BEA
Q_c=\frac{a\,T_c\,[\,\ln a\,]^2\,(n-1)\,\sqrt{1-\theta}}{[\,1+(n-1)a\,]^{2}},
\EEA
while the consumed work is
smaller and scales as $1-\theta$.

Eq.~(\ref{capo}) suggests that the maximization of $\chi$ 
imposes a lower bound on the efficiency: 
\BEA
\zeta>\zeta_{\rm CA}\equiv \frac{1}{\sqrt{1-\theta}}-1.
\EEA
This is numerically checked to be the case for all $0<\theta<1$ and all $n$; see also
Fig.~\ref{f2}. 

The expression of $\zeta_{\rm CA}$ was already obtained within
finite-time thermodynamics|but as an upper bound on the efficiency|and
argued to be an analogue of the Curzon-Ahlborn efficiency for
refrigerators \cite{yan_chen,velasco}.  Section \ref{class} explains
that also within the present microscopic approach $\zeta_{\rm CA}$ can
be an upper bound for $\zeta$ provided that $\chi$
is maximized under certain constraints. 

Recalling (\ref{peshavar}), our discussion after
(\ref{aharon1}--\ref{aharon3}) and (\ref{kandagar}), we can interpret
the lower bound for the efficiency as a lower bound on the intermediate
temperature $T'_c$ of $\S$:
\BEA
\frac{1-\sqrt{1-\theta}}{\theta}<\frac{T'_c}{T_c}<1, 
\label{stopsignal}
\EEA
i.e., the lowest temperature
$T_c'$ cannot be too low under optimal $\chi$. Compare this with
the fact that under vanishing efficiency (that is for very large amount of
the consumed work), $T_c'$ can be arbitrary low; see our discussion
after (\ref{peshavar}). Thus a well-defined lowest (per cycle)
temperature emerged once we restricted the resource of cooling (the consumed work). 

\section{The many-level regime: Reaching the Carnot limit at a finite power.}
\label{carnot_richting}

Now we turn to studying the regime 
\BEA
\ln (n-1)\gg 1. 
\label{marafet}
\EEA
First of all, let us introduce two new
variables
\BEA
\rho \equiv \u (n-1)\ln [n-1], ~~~~
\xi \equiv \frac{\ln [n-1]}{\v(n-1)},
\label{barsa}
\EEA
denote 
\BEA
\label{gabla}
p\equiv \ln [n-1], 
\EEA
and rewrite $\chi$ in (\ref{14}) as 
\BEA
\label{kokand}
\frac{\chi (1-\theta)}{\theta T_c p}= \frac{(1-\frac{\rho\xi}{p^2})(1+\frac{1}{p}\ln [\frac{\xi}{p}])^2}
{(1+\frac{\rho}{p})(1+\frac{\xi}{p})(1+\frac{1}{(1-\theta)p}\ln[\frac{p^{\theta+1}}{\rho \xi^\theta}]   )}.
\EEA
The expression in RHS of (\ref{kokand}) is now to be optimized over
$\rho$ and $\xi$.  We note that if these parameters stay finite in the
limit $p\equiv \ln [n-1]\gg 1$, the value of $\chi$ is read off
directly: $\frac{\chi (1-\theta)}{\theta T_c p}=1$.  The finitness of
$\rho$ and $\xi$ in the limit $p\equiv \ln [n-1]\gg 1$ is confirmed by
expanding the RHS of (\ref{kokand}) over the small parameter
$\frac{1}{p}$, collecting terms $\propto {\cal O}(\frac{1}{p})$,
differentiating them over $\rho$ and $\xi$, and equating the resulting
expressions to zero. This produces:
\BEA
\label{madrid}
\rho=\frac{1}{1-\theta}+{\cal O}\left(\frac{1}{p}\right), ~~~\xi=\frac{2-\theta}{1-\theta}+{\cal O}\left(\frac{1}{p}\right). 
\EEA
Substituting these into (\ref{peshavar}) and (\ref{14}) we get
\begin{gather}
\zeta=\frac{\theta}{1-\theta}-\frac{2\theta}{(1-\theta)^2}\frac{\ln[p]}{p}
\label{tacit}
+{\cal O}\left[ \frac{1}{p^2} \right],
\\
\frac{Q_c}{T_c}=p-\frac{3-\theta}{1-\theta}-\ln\left[ \frac{1-\theta}{2-\theta}\, p  \right]
+{\cal O}\left[ \frac{1}{p} \right].  
\label{tacit2}
\end{gather}
Note from (\ref{marafet}, \ref{gabla}) that the dominant factor in the
efficiency $\zeta$ is the Carnot value $\frac{\theta}{1-\theta}$, while
the subleading term is naturally negative; see (\ref{tacit}).  Likewise,
the dominant factor in $\frac{Q_c}{T_c}$ is $p\equiv \ln [n-1]$, while
the subleading term is ${\cal O}(1)$.  We also see that the limit $\ln
(n-1)\gg 1$ does not commute with the equilibrium limit $\theta\to 1$, since the
corrections in (\ref{tacit}, \ref{tacit2}) diverge for $\theta\to 1$. 

\begin{figure}
\includegraphics[width=8cm]{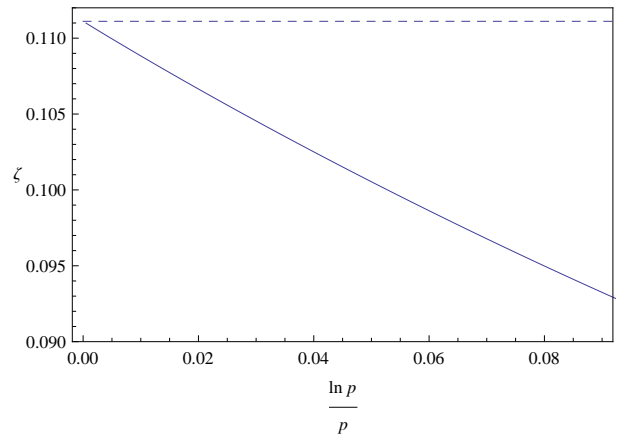}
\caption{Convergence of the efficiency $\zeta$ (normal line) to the Carnot value
$\eta_{\rm C}=1/9$ (dashed line) as a function of $\frac{\ln p}{p}$, where 
$p=\ln[n-1]$; see (\ref{marafet}--\ref{kokand}). } 
\label{f22}
\end{figure}

Thus in this regime $p\equiv\ln (n-1)\gg 1$ the efficiency converges to the
Carnot value; see Fig.~\ref{f22}.

Recalling (\ref{burundi}) we see from (\ref{barsa}, \ref{madrid}) that
in the regime (\ref{marafet}) the total occupation of the higher levels
of $\R$ is small, so that $\R$ is predominantly in its ground state
before applying the work-consuming external pulse.  In contrast, $\S$ is
more probably in one of its excited states. These facts are expected,
because $\S$ has to give up some energy, while $\R$ has to accept it. 

We note that this regime resembles in several aspects the macroscopic
regime $N\gg 1$ of a $N$-particle system. Recall that for $N\gg 1$
(weakly coupled) particles the number of energy levels scales as $e^N$,
while energy scales as $N$. Now for the above situation (\ref{tacit2})
the transferred heat $Q_c$ is (in the leading order) a product of the
colder temperature $T_c$ and the "number of particles" $\ln (n-1)$. 

The effective temperatures $T_h'$ and $T_c'$ [see (\ref{aharon1}--\ref{aharon3})]
in this limit are close to their initial values:
\BEA
\frac{T_h'}{T_h}=1+\frac{1}{p}\ln\left[\frac{(1-\theta)^2p^2}{2-\theta}  \right]+{\cal O}\left[ \frac{1}{p^2} \right], \\
\frac{T_c'}{T_c}=1-\frac{1}{p}\ln\left[\frac{(1-\theta)^2p^2}{2-\theta}  \right]+{\cal O}\left[ \frac{1}{p^2} \right], 
\EEA
where we employed (\ref{barsa}, \ref{madrid}). Though during the
refrigeration process the systems $\S$ and $\R$ are able to process
large amounts of work and heat ($\propto \ln [n-1]$), their temperatures
are not perturbed strongly.

\subsection{Finiteness of power}
\label{popo}

It is important to note that the asymptotic attainability (\ref{tacit})
of the Carnot bound for $\zeta$ is related to a finite transferred heat
$Q_c=T_c\ln [n-1]$, but it also can be related to a finite power
$\frac{Q_c}{\tau}$, where in our model the cycle time $\tau$ basically
coincides with the relaxation time; recall our discussion in section
\ref{po}.  This appears to be unexpected, because within the standard
thermodynamic analysis the Carnot efficiency is reached by the Carnot
cycle at a vanishing power \cite{callen}; see section \ref{carcar} for a
precise meaning of this statement. In any refrigerator model known to
us|see, e.g., \cite{segal}|approaching the Carnot limit
means nullifying the power. See also in this context our discussion
around equation (\ref{bekum}); various reasons preventing the approach
to the Carnot efficiency for thermal machines (even for small machines
working at zero power) are analysed in \cite{sekimoto}. Now we
supplement our discussion in section \ref{po} with more specific arguments.

We already stressed in section \ref{po} that within the second stage of
the refrigerator functioning, where both $\R$ and $\S$ relax to
equilibrium under influence of the corresponding thermal baths, the
relaxation mechanism can be associated with the collisional system-bath
interaction; see Appendix \ref{app_collisions} for a detailed discussion
of this mechanism. Here there are three characteristic
times: the single collision duration time $\tau_{\rm col}$ is much
smaller than the inter-collision time $\tau_{\rm btw}$, while the
relaxation of the system to its equilibrium state is governed by the
time $\tau_{\rm rel}$. The assumed condition $\tau_{\rm col}\ll\tau_{\rm btw}$ 
allows implementing the thermally isolated work-consuming pulse, because if the
pulse time is also much smaller than $\tau_{\rm btw}$, the pulse does not 
overlap with collisions. 

In Appendix \ref{app_collisions_1} we study the relaxation time of the
system with $n-1$ fold degenerate upper energy levels and
non-degenerate lowest energy level. We also account for the limit
$\ln[n-1]\gg 1$, where the Carnot efficiency is reached; see
(\ref{tacit}). It is shown that for such a system the relaxation time
$\tau_{\rm rel}$ can|depending on the details of the thermal bath and
its interaction with the system|range from few $\tau_{\rm btw}$'s to a
very long times $\propto n \tau_{\rm btw}$. The former relaxation time
means a finite power, while the latter time implies vanishing
power $\propto \frac{\ln n}{n}$ for $\ln n\gg 1$.

These two extreme cases are easy to describe without addressing the
formalism of Appendices \ref{app_collisions_0} and
\ref{app_collisions_1}. 

{\bf 1.} For simplicity let us focus on the relaxation of the system
$\R$ that after the work-extracting pulse (\ref{cobra}) is left in the
state $\sigma$, and is now subjected to a stream of the bath particles
(the situation with $\S$ is very similar).  Recall that each bath
particle before colliding with $\R$ is in the Gibbsian equilibrium state
at the temperature $T_h$.  Now assume that each bath particle also has
$n-1$ fold degenerate upper level, and one lowest energy level.  Also,
the non-zero energy spacing for the bath particles is equal to that of
$\R$.  Then the relaxation of $\R$ is achieved just after one collision
provided that the system-bath interaction [during this collision]
amounts to a SWAP operation. Note that the characteristic time of this
relaxation is $\tau_{\rm btw}$, and that this is a non-exponential
scenario of relaxation, because the system exactly settles into its
equilibrium Gibbsian state after the first collision. 

No work is done during collisional relaxation; see Appendix
\ref{app_collisions}.  Indeed, under above assumptions on the energy
levels of the bath particles, the SWAP operation commutes with the free
Hamiltonian $H_{1}\otimes 1+1\otimes H_\R$ (where $H_{1}$ is the
Hamiltonian of the given bath particle), which implies that the final
energy of $\R$ plus the bath particle is equal to its initial value.
Since each separate collision is a thermally isolated process, this
means that no work is done; see (\ref{work}). 

{\bf 2.}
If each collision is very weak and almost does not exchange heat with
the system $\R$, the relaxation time becomes very long.  Intermediate cases
are discussed in Appendix \ref{app_collisions_1}.  These intermediate
cases are relevant, since the power of refrigeration is finite even for
long relaxation times $\propto \ln n$. Indeed, we recall from
(\ref{tacit2}) that $Q_c=T_c \ln [n-1] +{\cal O}(1)$. 

\subsection{More realistic spectra still allowing to reach the Carnot bound }

One can ask whether the convergence (\ref{tacit}) to the Carnot bound is
a unique feature of the spectra (\ref{burundi}) in the limit $\ln
[n-1]\gg 1$, or whether there are other situations that still allow
$\zeta\to\zeta_{\rm C}$. The answer is positive as we now intend to
show. For the energy spectra (\ref{2star}) we postulate [$k=1,\ldots,
n-1$]
\BEA
\eps_{k+1}=\eps+(k-1)\delta, ~~ \mu_{k+1}=\mu+(k-1)\delta, 
\label{igla}
\EEA
where $\delta>0$ is a parameter. Next, we assume that the following six conditions hold
\begin{gather}
\label{bug}
(n-1)\beta\delta\gg 1, \quad \beta_c\delta\ll 1 , \quad \beta_h\delta\ll 1 \\ 
\label{bora}
\bar{p}\equiv \ln [{T_c}/{\delta}]\gg 1, \\ 
u\equiv e^{-\beta_h \eps}\simeq \delta/\bar{p}, \qquad
v\equiv e^{-\beta_c \mu} \simeq \bar{p}\,\delta.
\label{tel}
\end{gather}
Under conditions (\ref{bug}, \ref{bora}, \ref{tel})|and {\it assuming} the SWAP operation for the pulse|we show below 
that the results analogous to 
(\ref{tacit}, \ref{tacit2}) hold, 
\BEA
\zeta=\frac{\theta}{1-\theta}
+{\cal O}\left[ \frac{1}{\bar{p}} \right],~~~~
\frac{Q_c}{T_c}=\bar{p} \left(1+  {\cal O}\left[ \frac{1}{\bar{p}} \right] \right).
\label{toro}
\EEA
where the role of a large parameter $p=\ln [n-1]$ in (\ref{tacit},
\ref{tacit2}) is now played by $\bar{p}$ \footnote{In the definition
(\ref{bora}) of $\bar{p}$ one can as well employ $T_h$ instead of $T_c$.
This will not lead to serious changes, because we always assume that
$\theta= T_c/T_h$ is fixed. }. Note that (\ref{bug}) and (\ref{bora})
still imply that $\ln[n-1]\gg 1$. 

The spectra (\ref{igla}) under condition (\ref{bug}, \ref{bora},
\ref{tel}) correspond to a quasi-continuous part separated from the
ground state by a gap. This type of spectrum avoids the strong
degeneracy of (\ref{burundi}), and is met in conventional
superconductors below the transition temperature \cite{tinkham}. 

To derive (\ref{toro}) via (\ref{igla}--\ref{tel}) we note the following formulas for
the partition sums $Z_h\equiv{\tr\, [e^{-\beta_h H_\R}]}$ and $Z_c\equiv{\tr\, [e^{-\beta_c H_\S}]}$
in (\ref{1}), the heat $Q_c$ and work $W$:
\BEA
&&Z_h=1+\frac{T_h u}{\delta}, ~~~~
Z_c= 1+\frac{T_c v}{\delta},\nonumber\\
&&Q_c =T_c\ln\left[\frac{1}{v}\right] \left[\frac{1}{Z_h} - \frac{1}{Z_c}\right]+\frac{T_c^2 v}{Z_c\delta}-\frac{T_h^2 u}{Z_h\delta}, \nonumber\\
&&W= T_h \left[\ln\frac{1}{u}-\theta \ln\frac{1}{v}\right]\left[\frac{1}{Z_h} - \frac{1}{Z_c}\right].
\nonumber
\EEA

\section{ Entropy production}   
\label{entropy_production}

Entropy production is an important characteristics of thermal machines, because 
it quantifies the irreversibility of their functioning \cite{lindblad,callen}.
For our refrigerator model, no entropy is
produced during the first stage, which is thermally isolated from the
baths. However, a finite amount of entropy is produced during the
second, relaxational stage. The overall entropy production reads
\BEA
S_i=S[\Omega_{\rm fin}||\Omega_{\rm in}], \nonumber
\EEA
and controls the
deviation of efficiency from its maximal Carnot value; see
(\ref{laplace}). In the optimal conditions (\ref{cobra}, \ref{burundi}), we get
\BEA
S_i &=& S[\rho||\sigma]+S[\sigma||\rho] \\
&=&
\frac{\ln [\bar{v}/\bar{u}]\,(\bar{v}-\bar{u}) (n-1) }
{[\,1+(n-1)\,\bar{v}\,][\,1+(n-1)\,\bar{u}\,]}, 
\label{bratva}
\EEA
where $S[\sigma||\rho]$ and $S[\rho||\sigma]$ are the entropies produced
in, respectively, cold and hot bath.  Indeed, consider the system $\S$
that after the external field action is left in the state with density
matrix $\rho$ [see (\ref{cobra})], and now under influence of the
thermal bath should return to its initial state $\sigma\propto
e^{-\beta_c H_{\S}}$. 
Now
\BEA
T_c S[\rho||\sigma] 
={\rm tr}[H_\S  \rho ]-T_c S[\rho]+T_c\ln\,{\rm tr} \, e^{-\beta_c H_{\S}},
\label{beknazarov}
\EEA
is the difference between the non-equilibrium free energy ${\rm tr}[H_\S  \rho ]-T_c S[\rho]$ of $\S$ in the state $\rho$
[and in contact to a thermal bath at temperature $T_c$]
and the equilibrium free energy $-T_c\ln\,{\rm tr} \, e^{-\beta_c H_{\S}}$. 
Simultaneously, $T_c S[\rho||\sigma] $ in (\ref{beknazarov}) is the maximal work that can be extracted
from the system $\S$ (in state $\sigma$) in contact with the $T_c$-bath
\cite{armen}. During relaxation this potential work is let to relax into
the $T_c$-bath increasing its entropy by $S[\rho||\sigma]$. Likewise,
$S[\sigma||\rho]$ is the entropy production during the relaxation of
the system $\R$ in contact with the $T_h$-bath.

Now in the regime $\ln[n-1]\gg 1$, $S_i$ amounts to $\ln
[\bar{v}/\bar{u}]\simeq \ln(\ln [n-1])$, see (\ref{bratva}), while the
consumed work $W$ and the transferred heat $Q_c$ scale as $\ln [n-1]$.
In other words, the entropy production $S_i$ is much smaller than both
$W$ and $Q_c$. This explains why for a large $\ln [n-1]$ the Carnot
efficiency is reached; see (\ref{laplace}, \ref{tacit}). 

In the equilibrium limit $\theta\to 1$, $S_i$ reads
\BEA
S_i=\frac{a\,[\,\ln a\,]^2\,(n-1)\,(1-\theta)}{[\,1+(n-1)a\,]^{2}},
\label{borukhi}
\EEA
where $a$ is given by (\ref{golem}), and where in deriving
(\ref{borukhi}) we employed (\ref{bratva}) and asymptotic expansions
presented after (\ref{golem}). Note that now $S_i$ is smaller than
$Q_c\propto \sqrt{1-\theta}$, but has the same scale $1-\theta$ as the
consumed work $W$; see our discussion after (\ref{capo}). Thus $S_i$
cannot be neglected, and this explains why the Carnot efficiency is not
reached in the equilibrium limit $\theta\to 1$; see (\ref{laplace}).

\section{Classical limit.} 
\label{class}

We saw above that the optimization of the target quantity
$\chi=Q_c\zeta$ produced an inhomogeneous type of spectrum, where a
batch of (quasi)degenerate energy levels is separated from the ground
state by a gap. It is meaningful to carry out the optimization
of $\chi$ {\it imposing} a certain homogeneity in the spectra of $\R$
and $\S$. The simplest situation of this type is the {\it equidistant}
spectra
\BEA
\label{remo}
\eps_n=(n-1)\eps, ~~~~
\mu_n=(n-1)\mu, 
\EEA
for $\R$ and $\S$; recall (\ref{2star}). For
$n\to\infty$ and $\eps\to 0$, $\mu\to 0$ these spectra correspond to the
classical limit. 

Thus, now we maximize $\chi=Q_c\zeta$ imposing conditions (\ref{remo}).
We found numerically that the optimal ${\cal U}$ again corresponds
to SWAP operation; see (\ref{cobra}). For $\chi=\chi(\u,\v)$ we get
\begin{gather}
\chi= 
\frac{T_c\theta \ln^2\frac{1}{\v}}
{\ln\frac{1}{\u}-\theta\ln\frac{1}{\v}}\left[
\frac{\v-\u}{(1-\v)(1-\u)}-\frac{n(\v^n-\u^n)}{(1-\v^n)(1-\u^n)}
\right],\nonumber
\end{gather}
where $\v=e^{-\beta_c\bar{\eps}}$ and $\u=e^{-\beta_h\bar{\mu}}$ are
found from maximizing $\chi$. The efficiency $\zeta$ is still given by
(\ref{11}). 

In the limit $n\gg 1$ we get from maximizing $\chi$:
\BEA
\bar{u}\to 1, ~~~~ \bar{v}\to 1~~~~ {\rm and}~~~~
\frac{n(\v^n-\u^n)}{(1-\v^n)(1-\u^n)}\to 0, 
\label{bonn1}
\EEA
implying that $\chi$ and $\zeta$
depend on one parameter $\phi\equiv\frac{1-\bar{u}}{1-\bar{v}}$: 
\BEA
\chi = \frac{T_c\theta (\phi-1)}{\phi(\phi-\theta)}, ~~~~~
\zeta = \frac{\theta}{\phi-\theta}.\nonumber
\EEA
The
optimal value of this parameter is $\phi=1+\sqrt{1-\theta}$. This leads to
\BEA
\chi=\frac{T_c\theta}{(1+\sqrt{1-\theta})^2}, ~~~
\zeta=\zeta_{\rm CA}=\frac{1}{\sqrt{1-\theta}}-1.  
\label{gomes}
\EEA
Thus for a large number of
equidistant energy levels the maximization of $\chi$ leads to
homogeneity ($\bar{\eps}\to 0$, $\bar{\mu}\to 0$), which is an
indication of the classical limit. The efficiency $\zeta$ in this
constrained optimal situation is equal to $\zeta_{\rm
CA}$.

The above results refer to optimizing $\chi$ in the limit $n\gg 1$. However,
we confirmed numerically that the above values (\ref{gomes}) for $\chi$
and $\zeta$|obtained in the limit $n\gg 1$|are upper bounds for $\chi$
and $\zeta$ at a finite $n$. 

Our conclusion is that the efficiency $\zeta_{\rm CA}$|which is a lower
bound for the efficiency during the unconstrained optimization of
$\chi$|appears to be an upper bound for refrigerators that operate under
equidistant (classical) spectra. The upper bound is reached in the limit $n\gg 1$.

These facts clarify to some extent why $\zeta_{\rm CA}$ was obtained as
an upper bound for the efficiency within the finite-time thermodynamics
(FTT) \cite{yan_chen,velasco}. Apparently, the quasi-equilibrium assumptions of
FTT implied constraints which are equivalent to imposing the homogeneous
spectra in our approach.

\section{Summary}
\label{summa_contra_gentiles}

We have studied a model of a refrigerator aiming to
understand its optimal performance at a finite cooling power; see
Fig.~\ref{f1}.  The structure of the model is such that it can be
optimized over almost all its parameters; additional constraints can and
have been considered, though. We have confirmed the complementarity
between optimizing the heat $Q_c$ transferred from the cold bath $T_c$
and efficiency $\zeta$: maximizing one nullifies the other.  Similar
effect for different models of quantum refrigerators are reported
in \cite{feldman,kosloff_minimal_temperature,segal}. 

To get a balance between $Q_c$ and $\zeta$ we have thus chosen to optimize
their product $\zeta Q_c$. This leads to a {\it lower} bound $\zeta_{\rm
CA}=\frac{1}{\sqrt{1-\theta}}-1$ ($\theta\equiv \frac{T_c}{T_h}$) for
the efficiency in addition to the upper Carnot bound $\zeta_{\rm
C}=\frac{1}{1-\theta}-1$. The fact of $\zeta>\zeta_{\rm CA}$ implies that there is
the lowest finite temperature reachable within one cycle of refrigeration; see 
(\ref{stopsignal}).

The lower bound $\zeta_{\rm CA}$ is reached in the equilibrium limit
$T_c\to T_h$. Constraining both systems to have homogeneous (classical)
spectra, $\zeta_{\rm CA}$ is reached as an upper bound. This is just
like within finite-time thermodynamics (FTT), when maximizing the
product of the cooling-power and efficiency \cite{yan_chen}, or the
ratio of the efficiency and the cycle time \cite{velasco}.  In this
sense $\zeta_{\rm CA}$ seems to be universal. It may play the same role
as the Curzon-Ahlborn efficiency for heat engines $\eta_{\rm CA}$,
which, again, is an upper bound within FTT \cite{ca}, but appears as a
lower bound for the engine models studied in
\cite{armen}. For other opinions on the Curzon-Ahlborn
efficiency for refrigerators see \cite{jimenez,unified}. 

The Carnot upper bound is asymptotically reached in the many-level limit
of the model.  We saw that this asymptotic convergence is related to a
finite heat transferred per cycle, and we argued that it can also be
related to a finite power if the relaxation scenario of the model
refrigerator is chosen properly: provided that the cycle time is larger than
the relaxation time, one can perform exponentially large number of
refrigeration cycles before {\it inevitable} deviations from
cyclicity|that in any case are there due to a finite cycle time|will
accumulate.  To our knowledge such an effect has never been seen so far
for refrigerator models.

For the optimal refrigerator the transferred heat $Q_c$ behaves as
$Q_c\propto T_c$ (in particular, for $T_c\to 0$); see (\ref{10},
\ref{14}, \ref{tacit}). This is in agreement with the optimal
low-temperature behaviour of $Q_c$ from the viewpoint of the third law
\cite{kosloff_jap,rezek}. 

\subsection*{Acknowledgement}
This work has been supported by Volkswagenstiftung.

We thank K. G. Petrosyan for a useful suggestion.

\appendix

\section{Collisional relaxation.}
\label{app_collisions}

\subsection{General consideration}
\label{app_collisions_0}

The purpose of this discussion is to outline the general structure of a
collisional relaxation process. Our presentation follows to
\cite{partovi,mityugov}. 

The thermal bath is modeled as a collection of $N\gg 1$ independent
equilibrium systems (particles) with initial density matrices
$\omega_{i}=\frac{1}{Z_i}\exp[-\beta H_{i}]$ and Hamiltonians $H_{i}$,
where $i=1,..,N$, and where $1/\beta=T$ is the bath temperature. This
formalizes the intuitive notion of the bath as a collection of many
weakly-interacting particles. 

The target system $\R$ starts in [an arbitrary] initial state $\rho_\R$
and has Hamiltonian $H_{\R}$. The collisional relaxation is realized
when the particles of the bath sequentially interact [collide] with
$\R$. Multiple collisions (between the target system and simultaneously
two or more bath particles) are neglected. 

Consider the first 
collision. The initial state of $\R$ and the first bath particle is
$\Omega_{1+\R}=\rho\otimes\omega_1$. The interaction between them is realized via a unitary
operator ${\cal V}$, so that the final state after the first collision is 
$\Omega_{1+\R}'={\cal V}\Omega_{1+\R}{\cal V}^\dagger$. 
This unitary operator is generated by the full Hamiltonian $H_{1+\R}$:
\BEA
\label{abner}
H_{1+\R} = H_{1}+H_{\R}+H_{1\,,\,\R},
\EEA
where $H_{1\,,\,\R}$ is the interaction Hamiltonian.
Define separate final states:
\BEA
\rho'={\rm tr}_1 \Omega_{1+\R}', \qquad
\omega_1'={\rm tr}_\R \Omega_{1+\R}',
\EEA
where ${\rm tr}_1$ and ${\rm tr}_\R$ are the partial over the first particle and $\R$, respectively.
Recall the definition (\ref{43}) of the relative
entropy. The unitarity of ${\cal V}$ implies
\BEA
S[\,\Omega_{1+\R}'\, ||\, \rho'\otimes \omega_1\, ]&=&
{\rm tr}[\Omega_{1+\R}\ln \Omega_{1+\R}
]\nonumber\\
&-&{\rm tr}[\Omega_{1+\R}'\ln(\rho'\otimes \omega_1)].
\label{ortega}
\EEA
Employing $\omega_{1}=\frac{1}{Z_1}\exp[-\beta H_{1}]$ and
$S[\,\Omega_{1+\R}'\, ||\, \rho'\otimes \omega_1\,]\geq 0$ 
in (\ref{ortega}) we get
\BEA
\label{kabanets}
T \Delta S_{\R}+\Delta U_1\geq 0,
\EEA
where $\Delta S_{\R}={\rm tr}\left [-\rho'\ln \rho' + \rho\ln \rho 
\right]
$ and $\Delta U_{1}={\rm tr}(H_1\left  [\omega_1'-\omega_1\right])$
are, respectively, the change of the entropy of R and the average energy of 
the first particle. 

We now require that the interaction ${\cal V}$ conserves the {\it average}
energy: 
\BEA
\label{urdu}
\Delta U_1=-\Delta U_R. 
\EEA
Using this in (\ref{kabanets}) one has
\BEA
\label{kabanets1}
\Delta U_\R -T \Delta S_{\R}\leq 0. 
\EEA
Since we did not use any special feature of the initial state of $\R$,
(\ref{kabanets1}) holds for subsequent collisions of $\R$ with the bath
particles. Thus $U_\R -T  S_{\R}$ decays in time, and it
should attain its minimum. It is well-known \cite{partovi,mityugov} that
this minimum is reached for the Gibbs matrix $\rho\propto e^{-\beta
H_\R}$: collisions can drive $\R$ to equilibrium starting from an
arbitrary state \cite{partovi,mityugov}. 

Condition (\ref{urdu}) expresses the average energy conservation. It is
natural to use a more stringent condition according to which the sum of
energies of $\R$ and the bath particle $1$ is conserved in time
\cite{mityugov}:
\BEA
\label{ortiz}
[H_{1}+H_{\R},H_{1\,,\,\R}]=0.
\EEA
This condition makes the dynamics {\it autonomous}, since for any
initial state of $\R+1$ the switching the interaction on and off does
not cost energy and (\ref{urdu}) holds automatically. 

For condition (\ref{ortiz}) to be non-trivial, the operator
$H_{1}+H_{\R}$ should have a degenerate spectrum. Otherwise due to
$[H_{1}+H_{\R},H_{1}]=0$ and (\ref{ortiz}), $H_{\R}$ and $H_1$ will be
constants of motion, which means that no transfer of energy and thus no
relaxation is possible. 

Here are two crucial points of the collisional relaxation.

{\bf 1.} If the target system starts in the equilibrium state, this
state does not change in time under subsequent collisions.  This
analogue of the zero law of thermodynamics is especially obvious from
condition (\ref{ortiz}), but it also holds simply from the conservation
of the average energy (\ref{urdu}). Indeed, if $\rho_{\bf H}$ is the
equilibrium state, $\rho_{\bf H}\propto e^{-\beta H_{\bf H}}$, and also
condition (\ref{urdu}) holds, the relative entropy $S[\,\Omega_{1+\R}'\,
||\, \rho_{\bf H}\otimes \omega_1]$ is equal to zero, which can happen
only for $\Omega_{1+\R}' =\rho_{\bf H}\otimes \omega_1$. 

{\bf 2.} No work is done for switching collisions on and off. This is
clearly seen from (\ref{ortiz}), which states that the free Hamiltonian
is a constant of motion. 

\subsection{The relaxation time for a pertinent example}
\label{app_collisions_1}

Now we study the relaxation time for an $n$-level system $\R$ under
collisional dynamics. We assume that $n-1$ levels of $\R$ coincide and
have energy $\eps>0$. The lowest energy level is not degenerate and has
energy zero. Importantly, we assume that condition (\ref{ortiz}) holds
meaning that the relaxation proceeds autonomously, i.e., without
additional energy [work] costs on switching the interaction on and off. The
initial (before colliding with the first bath particle) density matrix
of $\R$ is assumed to be Gibbsian at temperature $T_0=1/\beta_0$:
\BEA
\rho=\frac{e^{-\beta_0 H_\R}}{Z_\R(\beta_0)}
=r \P_0 +\frac{1-r}{n-1}\P_\eps, \quad r =\frac{1}{1+(n-1) e^{-\beta_0\eps}},\nonumber
\EEA
where $\P_0=|0\rangle\langle 0|$ and $\P_\eps$ is the projectors 
on the $n-1$-dimensional eigen-space of $\rho$ with eigenvalue $\eps$. 

To satisfy the degeneracy of the interaction Hamiltonian [see our
discussion after (\ref{ortiz})] we assume that the first bath particle
has (among others energies) energy levels $E$ and $E+\eps$. The
degeneracies of these levels are $n^{[1]}_{E}$ and $n^{[1]}_{E+\eps}$, respectively.
The equilibrium density matrix of the bath particle $1$ is written as
\BEA
\label{fru}
\rho_1=\widetilde{\rho}_1 +r^{[1]}_E \, \P_E^{[1]} +r^{[1]}_{\eps+E}\, \P_{\eps+E}^{[1]}, \\
r^{[1]}_{E_\alpha}={e^{-\beta E_\alpha}}/{Z_1}, \quad Z_1={\sum}_\alpha n^{[1]}_{E_\alpha} e^{-\beta E_\alpha},
\label{blob}
\EEA
where $r_E^{[1]}$ and $r_{E+\eps}^{[1]}$ are the Boltzmann weights for
the energy levels $E$ and $E+\eps$, respectively, the summation in
(\ref{blob}) is taken over all energy levels of the bath particle,
$\P_E^{[1]}$ and $\P_{\eps+E}^{[1]} $ are the projectors on the
corresponding sub-spaces,
\BEA
{\rm tr}\,\P_E^{[1]}=n^{[1]}_{E},    \qquad
{\rm tr}\,\P_{\eps+E}^{[1]}=n^{[1]}_{E+\eps}, 
\EEA
and where $\widetilde{\rho}_1$ in (\ref{fru}) is the remainder of $\rho_1$. 

It is assumed that the unitary operator ${\cal V}$ responsible for the
interaction operates within the sub-space with the projector
$\P_\eps\otimes \P_E^{[1]} + \P_0\otimes \P_{E+\eps}^{[1]}$ (this
sub-space has energy $E+\eps$), i.e.,
\BEA
[{\cal V},\P_\eps\otimes \P_E^{[1]} + \P_0\otimes \P_{E+\eps}^{[1]} ]=0.
\label{brams}
\EEA
Then the post-collision density matrix $\rho'$ of $\R$ reads
\begin{gather}
\rho'={\rm tr}_1{\cal V} \rho\otimes\rho_1{\cal V}^\dagger=\rho\nonumber\\
-
\left(r\,r^{[1]}_{E+\eps}-r^{[1]}_E\,\frac{1-r}{n-1}\right)[n^{[1]}_{E+\eps}\P_0
-{\rm tr}_1{\cal V}\P_0\otimes \P_{E+\eps}^{[1]} {\cal V}^\dagger]\nonumber
\end{gather}
Clearly, $\rho'$ commutes with $H_\R$. For simplicity, we choose ${\cal
V}$ such that the degeneracy of $\rho$ is not resolved, i.e., in the
state $\rho'$, the occupations of the higher energy levels of $\R$ are
equal.  This means we need to keep track of the lowest energy-level
occupation $\langle 0|\rho'|0\rangle\equiv r'$ only:
\begin{gather}
\label{cuba}
r'-r=-A\left [ r- r_{\rm eq} \right], \quad r_{\rm eq}\equiv \frac{1}{1+(n-1)e^{-\beta\eps}},\\
\label{gvadalajara}
A\equiv \frac{r_E^{[1]}}{r_{\rm eq}(n-1)}
\left [n^{[1]}_{E+\eps}
-\langle 0|\,(\, {\rm tr}_1{\cal V}\P_0\otimes \P_{E+\eps}^{[1]} {\cal V}^\dagger\,)|0\rangle \right],
\end{gather}
where $r_{\rm eq}$ is the equilibrium value of $r$.
$A$ can be maximized over the unitary ${\cal V}$ [under condition (\ref{brams})]
\BEA
\label{blumkin}
A_{\rm max}= \frac{r_E^{[1]}{\rm min}\left[\, n^{[1]}_{E+\eps}, n^{[1]}_{E}(n-1)\,\right]}{r_{\rm eq}(n-1)}. 
\EEA
Using (\ref{blob}) one can show that $A\leq A_{\rm max}\leq 1$: after first collision $\R$ gets closer to its equilibrium state;
see (\ref{cuba}). This equation obviously generalizes to subsequent collisions
[we revert from (\ref{cuba1}) to (\ref{cuba}) for $m=1$]:
\BEA
\label{cuba1}
r'_{[m]}-r'_{[m-1]}=(1-A)^m\left [ r- r_{\rm eq} \right], 
\EEA
It is seen that
(\ref{cuba1}) predicts exponential (with respect to the number of
collisions) relaxation towards the equilibrium value $r_{\rm eq}$ of
$r$. The approach to equilibrium is governed by the factor $(1-A)^n$
meaning that when $|A|\ll 1$ the effective number of collisions after
which the equilibrium is established equals to $-1/[\ln(1-A)]$. 

Now the shortest relaxation corresponds to just one collision and it is
reached for $A=1$, e.g., $r_E^{[1]}=r_{\rm eq}$ and
$n^{[1]}_{E+\eps}=n-1$ in (\ref{blumkin}). Then the corresponding
unitary operator ${\cal V}$ is the SWAP operation. The relaxation time
in this case amounts to one inter-collision time. 

It should be clear that there is no upper limit on the relaxation time.
The latter can be arbitrary large, e.g., due to ${\cal V}$ converging to
$1$ in (\ref{gvadalajara}). Various intermediate cases can be studied
with help of (\ref{blumkin}). In particular, it is not difficult to
identify regimes, where the relaxation time scales as $\propto \ln n$. 


\begin{thebibliography}{99}


\bibitem{lindblad}
G. Lindblad, {\it Non-Equilibrium Entropy and Irreversibility}
(Reidel, Dordrecht, 1983).

\bibitem{callen}
H.B. Callen, {\it Thermodynamics} (John Wiley, NY, 1985). 

\bibitem{landau}L.D. Landau and E.M. Lifshitz, {\it Statistical
Physics, I}, Pergamon Press Oxford, 1978.

\bibitem{q_t}J. Gemmer, M. Michel and G. Mahler, {\it Quantum Thermodynamics}
(Springer, 2004).


\bibitem{venturi}
V. Balzani, A. Credi and M. Venturi, {\it Molecular Devices and Machines} (Wiley-VCH, Weinheim, 2003). 

\bibitem{ftt}
R.S. Berry, V.A. Kazakov, S. Sieniutycz, Z. Szwast, and A.M. Tsvilin,
{\it Thermodynamic Optimization of Finite--Time Processes} (John Wiley
\& Sons, Chichester, 2000). 


\bibitem{ca}F. Curzon and B. Ahlborn, Am. J. Phys. {\bf 43}, 22 (1975).



\bibitem{armen}
A.E. Allahverdyan {\it et al.},
Phys. Rev. E {\bf 77}, 041118 (2008).


\bibitem{tu}Z.C. Tu, J. Phys. A {\bf 41}, 312003 (2008).

\bibitem{izumida_okuda}
Y. Izumida and K. Okuda, EPL {\bf 83}, 60003 (2008).

\bibitem{udo}
T. Schmiedl and U. Seifert, EPL {\bf 81}, 20003 (2008).

\bibitem{esposito}
M. Esposito {\it et al.}, Phys. Rev. Lett. {\bf 102}, 130602 (2009).

M. Esposito {\it et al.}, {\it Quantum-dot Carnot engine at maximum power}, arXiv:1001.2192.

\bibitem{jmod}A.E. Allahverdyan, R. Balian and Th.M. Nieuwenhuizen, 
J. Mod. Opt. {\bf 51}, 2703 (2004).


\bibitem{domi}D. Janzing, J. Stat. Phys. {\bf 122}, 531 (2006).

\bibitem{henrich}
M.J. Henrich {\it et al.}, Europhys. Lett., {\bf 76}, 1057 (2006).


\bibitem{kosloff_jap}R. Kosloff, E. Geva and J. M. Gordon, J. Appl. Phys. {\bf 87}, 8093 (2000).

\bibitem{feldman}T. Feldman and R. Kosloff, Phys. Rev. E, {\bf 61}, 4774 (2000).

\bibitem{kosloff_minimal_temperature}T. Feldmann and R. Kosloff, {\it 
Optimal Performance of Reciprocating Quantum Refrigerators}, arXiv:0906.0986.


\bibitem{segal}D. Segal, Phys. Rev. Lett. {\bf 101}, 260601 (2008).

\bibitem{rezek}Y. Rezek {\it et al.}, EPL, {\bf 85}, 30008 (2009).


\bibitem{yan_chen}Z. Yan and J. Chen, J. Phys. D {\bf 23}, 136 (1990).

\bibitem{velasco}S. Velasco {\it et al.}, Phys. Rev. Lett. {\bf 78}, 3241 (1997).

\bibitem{jimenez}B. Jimenez de Cisneros {\it et al.}, Phys. Rev. E {\bf 73}, 057103 (2006).

\bibitem{unified}
A. Calvo Hernandez {\it et al.}, Phys. Rev. E {\bf 63}, 037102 (2001).


\bibitem{dhar}R. Marathe, A. M. Jayannavar and A. Dhar, Phys. Rev. E {\bf 75}, 030103 (R) (2007).

D. Segal and A. Nitzan, Phys. Rev. E {\bf 73}, 026109 (2006).

J. Ren and B. Li, Phys. Rev. E {\bf 81}, 021111 (2010).


\bibitem{sekimoto}
K. Sekimoto, F. Takagi and T. Hondou, Phys. Rev. E {\bf 62}, 7759 (2000).

\bibitem{sekimoto_sasa}
K. Sekimoto and S. Sasa, J. Phys. Soc. Jpn. {\bf 66}, 3326 (1997).

\bibitem{minima} A. E. Allahverdyan and Th. M. Nieuwenhuizen, Phys. Rev. E {\bf 71}, 046107 (2005).

\bibitem{grif}R. B. Griffiths, J. Math. Phys. {\bf 6}, 1447 (1965).

\bibitem{scully}M. O. Scully {\it et al.}, J. Mod. Opt. {\bf 49}, 2297 (2002).


\bibitem{galindo}
A. Galindo and M. A. Martin-Delgado, Rev. Mod. Phys. {\bf 74}, 347 (2002).

\bibitem{swap}
N. Sangouard {\it et al.} Phys. Rev. A {\bf 72}, 062309 (2005).

\bibitem{klein}M. J. Klein, in {\it Thermodynamics of Irreversible Processes, 
Scuola internazionale di fisica "Enrico Fermi"}, ed. by S. R. de Groot (Bologna, 1960).

\bibitem{olkin}
A.W. Marshall and I. Olkin, {\it Inequalities: Theory
of Majorization and its Applications}, (Academic Press, New York,
1979).

\bibitem{partovi}
H. M. Partovi, Phys. Lett. A {\bf 137}, 440 (1989).

\bibitem{mityugov}
A. B. Brailovskii, V. L. Vaks, and V. V. Mityugov, Phys. Usp.
{\bf 166}, 795 (1996).

V. V. Mityugov, Phys. Usp. {\bf 170}, 681 (2000).


\bibitem{ruben}
A.E. Allahverdyan {\it et al.}, Phys. Rev. E {\bf 71}, 046106 (2005).

\bibitem{kms} R. Kubo, J. Phys. Soc. Jap. {\bf 12}, 570 (1957).

P. C. Martin and J. Schwinger, Phys. Rev. {\bf 115}, 1342 (1959).


\bibitem{tinkham}M. Tinkham, {\it Introduction to Superconductivity} (McGraw-Hill, 1975).

\end{thebibliography}
\end{document}